\documentclass[pre,showpacs,floatfix]{revtex4}
\usepackage{amssymb}
\usepackage{graphicx}

\begin{document}

\title{Interface solitons in locally linked two-dimensional lattices}
\author{ M. D. Petrovi\'c$^1$, G. Gligori\'c$^{1}$, A. Maluckov$^2$, Lj.
Had\v zievski$^1$, and B. A. Malomed$^3$}
\affiliation{$^1$ Vin\v ca Institute of Nuclear Sciences, University of Belgrade, P. O.
B. 522,11001 Belgrade, Serbia \\
$^2$ Faculty of Sciences and Mathematics, University of Ni\v s, P. O. B.
224, 18000 Ni\v s, Serbia\\
$^3$ Department of Physical Electronics, School of Electrical Engineering,
Faculty of Engineering, Tel Aviv University, Tel Aviv 69978, Israel}

\begin{abstract}
Existence, stability and dynamics of soliton complexes, centered at the site
of a single transverse link connecting two parallel 2D (two-dimensional)
lattices, are investigated. The system with the on-site cubic self-focusing
nonlinearity is modeled by the pair of discrete nonlinear Schr\"{o}dinger
equations linearly coupled at the single site. Symmetric, antisymmetric and
asymmetric complexes are constructed by means of the variational
approximation (VA) and numerical methods. The VA demonstrates that the
antisymmetric soliton complexes exist in the entire parameter space, while
the symmetric and asymmetric modes can be found below a critical value of
the coupling parameter. Numerical results confirm these predictions. The
symmetric complexes are destabilized via a supercritical symmetry-breaking
pitchfork bifurcation, which gives rise to stable asymmetric modes. The
antisymmetric complexes are subject to oscillatory and exponentially
instabilities in narrow parametric regions. In bistability areas, stable
antisymmetric solitons coexist with either symmetric or asymmetric ones.
\end{abstract}

\maketitle

\section{Introduction}

Solitons trapped at interfaces between different nonlinear media \cite%
{surface-theory}, \cite{surface-experiment}, or pinned by defects \cite%
{3,defect-theory,molinakivshar,defect-experiment,Roberto,novi}, have been
the subject of many recent studies. It has been found that the self-trapped
surface modes possess novel properties in comparison with the solitons in
bulk media. Among noteworthy features of these localized modes are a
threshold value of the norm, above which they exist, and the coexistence of
different surface modes with equal norms.

The studies of interface solitons in discrete systems have shown that
spatially localized states with broken symmetries can exist \cite%
{molinakivshar}, which may be related to the general phenomenon of the
spontaneous symmetry breaking (SSB) in bimodal nonlinear symmetric settings
with a linear coupling between two subsystems. For the first time, the SSB
bifurcation, which destabilizes symmetric states and gives rise to
asymmetric ones, was predicted in a discrete self-trapping model in Ref.
\cite{Scott}. This finding was followed by the prediction of the SSB in the
model of dual-core nonlinear optical fibers \cite{Wright}, \cite{Snyder}. In
the framework of the nonlinear Schr\"{o}dinger (NLS) equation, the concept
of the SSB was, as a matter of fact, first put forward in early work \cite%
{earlier}. Related to this context is the analysis of the SSB of discrete
solitons in the system of linearly coupled 1D and 2D (one- and
two-dimensional) discrete nonlinear-Schr\"{o}dinger (DNLS)\ equations \cite%
{VA} (the general outline of the topic of DNLS equations was given in book
\cite{PGK}).

The SSB was also analyzed in detail for solitons in the continual model of
dual-core fibers with the cubic (Kerr) nonlinearity \cite{Akhmediev}-\cite%
{Progress}, and in related models of Bose-Einstein condensates (BEC's)
loaded into a pair of parallel-coupled cigar-shaped traps \cite{1D-BEC}. In
the latter context, the analysis was generalized for 2D coupled systems \cite%
{2D-BEC}. The SSB for gap solitons was studied too, in the model of
dual-core fiber Bragg gratings \cite{Mak}, and later in the model of the
BEC\ trapped in the dual-trough potential structure, combined with a
longitudinal periodic potential \cite{Warsaw}. As concerns the relation
between discrete and continual systems, it is relevant to mention work \cite%
{Dong}, where exact analytical solutions were found for the SSB in the model
with the nonlinear coefficient in the form of a pair of delta-functions
embedded into a linear medium. In its own turn, the latter model has its own
discrete counterparts, in the form of a pair of nonlinear sites embedded
into a linear chain \cite{Molina,Valera}, or side-coupled to its \cite%
{Almas}. These models may be realized in terms of BEC and optics alike. The
SSB in such settings was recently analyzed in Refs. \cite{Valera} and \cite%
{Almas}, respectively.

Coming back to discrete media, the objective of the present work is to study
localized modes at the interface of two 2D uniform lattices with the cubic
on-site self-focusing, which are linearly linked at a single site. This link
plays the role of the interface. Accordingly, the localized modes are
complexes formed by two fundamental solitons in each lattice, centered at
the linkage site. We focus on the symmetry of the soliton complexes, with
the intention to investigate the SSB transitions in them.

This work is a natural extension of the recent study of interface modes in
the system of single-site-coupled 1D nonlinear lattices \cite{reshetke}.
Unlike the 1D situation, it would be difficult to realize such a 2D setting
in optics. However, it is quite possible in BEC: one may consider two
parallel pancake-shaped traps combined with a deep 2D\ optical lattice
traversing both pancakes \cite{Bloch}, with the local link induced by a
perpendicular narrow laser beam. In fact, similar two-tier layers of
nonlinear oscillators, transversely linked at sparse sites, can be realized
in a number of artificially built systems.

The article is organized as follows. The model is formulated in Section 2.
In the same section the existence and stability of various on-site-centered
fundamental localized modes are considered in a quasi-analytical form by
means of the variational approximation (VA) and Vakhitov-Kolokolov (VK)
stability criterion. Numerical results for the soliton complexes are
presented in Section 3, including the stability and dynamics. The numerical
findings are compared to the predictions of the VA, and both the analytical
and numerical results are compared to those reported in Ref. \cite{reshetke}
for the fundamental localized modes in the 1D counterpart of the system
with parallel-coupled lattices, or 'system 2', in terms of Ref.
\cite{reshetke}). The paper is concluded by Section 4.

\section{The model and variational approximation}

\subsection{The formulation}

The set of the locally linked 2D uniform lattices is displayed in Fig. \ref%
{fig1}. The intra-site coupling constant in the lattices is $C>0$, and $%
\varepsilon >0$ is the strength of the transverse link. The lattice system
is modeled by the following DNLS system,%
\begin{eqnarray}
i\frac{d\phi _{n,m}}{dt}+\frac{C}{2}(\phi _{n+1,m}+\phi _{n-1,m}+\phi
_{n,m+1}+\phi _{n,m-1}-4\phi _{n,m})+\varepsilon \psi _{n,m}\delta
_{n,0}\delta _{m,0}+\gamma \left\vert \phi _{n,m}\right\vert ^{2}\phi _{n,m}
&=&0,  \nonumber \\
i\frac{d\psi _{n,m}}{dt}+\frac{C}{2}(\psi _{n+1,m}+\psi _{n-1,m}+\psi
_{n,m+1}+\psi _{n,m-1}-4\psi _{n,m})+\varepsilon \phi _{n,m}\delta
_{n,0}\delta _{m,0}+\gamma \left\vert \psi _{n,m}\right\vert ^{2}\psi _{n,m}
&=&0,  \label{eq1}
\end{eqnarray}%
where $\gamma >0$ is the coefficient of the on-site self-focusing
nonlinearity, $t$ is the time, and $\delta _{m,n}$ is the Kronecker's
symbol. By means of obvious rescaling, we set $C/2=\gamma =1$.

\begin{figure}[ht]
\center\includegraphics [width=8cm]{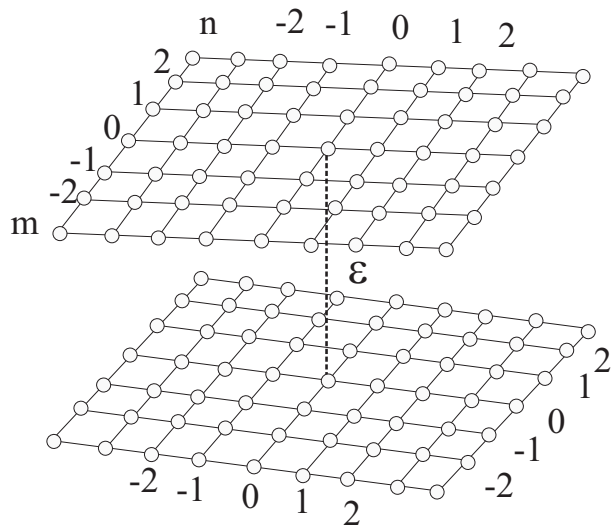}
\caption{(Color online) A schematic presentation of parallel 2D identical
latices, linearly linked at the single site $n=m=0$, with the coupling
constant $\protect\varepsilon $.}
\label{fig1}
\end{figure}

To construct soliton complexes formed around the transverse link, we look
for stationary solutions, $\phi _{n,m}=u_{n,m}\exp \left( -i\mu t\right) $, $%
\psi _{n,m}=v_{n,m}\exp \left( -i\mu t\right) $, where $u_{n,m},v_{n,m}$ and
$\mu $ are real lattice fields and the propagation constant, respectively.
The corresponding stationary equations following from Eqs. (\ref{eq1}) are
\begin{eqnarray}
\mu u_{n,m}+(u_{n+1,m}+u_{n-1,m}+u_{n,m+1}+u_{n,m-1}-4u_{n,m})+\varepsilon
v_{n,m}\delta _{n,0}\delta _{m,0}+\left\vert u_{n,m}\right\vert ^{2}u_{n,m}
&=&0,  \nonumber \\
\mu v_{n,m}+(v_{n+1,m}+v_{n-1,m}+v_{n,m+1}+v_{n,m-1}-4v_{n,m})+\varepsilon
u_{n,m}\delta _{n,0}\delta _{m,0}+\left\vert v_{n,m}\right\vert ^{2}v_{n,m}
&=&0.  \label{eq2}
\end{eqnarray}%
Equations (\ref{eq2}) can be derived from the Lagrangian,%
\begin{equation}
L=L_{u}+L_{v}+2\varepsilon u_{0,0}v_{0,0},  \label{eq3}
\end{equation}

\begin{eqnarray}
L_{u} &\equiv &\sum_{n=-\infty }^{+\infty }\sum_{m=-\infty }^{+\infty
}\left( (\mu -4)u_{n,m}^{2}+\frac{1}{2}%
u_{n,m}^{4}+2u_{n,m}(u_{n+1,m}+u_{n,m+1})\right) ,  \nonumber \\
L_{v} &\equiv &\sum_{n=-\infty }^{+\infty }\sum_{m=-\infty }^{+\infty
}\left( (\mu -4)v_{n,m}^{2}+\frac{1}{2}%
v_{n,m}^{4}+2v_{n,m}(v_{n+1,m}+v_{n,m+1})\right) ,  \label{eq4}
\end{eqnarray}%
where $L_{u}$ and $L_{v}$ are the intrinsic Lagrangians of the uncoupled
lattices, and the last term in Eq. (\ref{eq3}) accounts for the coupling
between them.

\subsection{The variational approximation}

The variational method follows the route described in Ref. \cite{VA}. We
adopt a natural ansatz, which was first applied to discrete lattices in Ref.
\cite{Weinstein}:%
\begin{equation}
\left\{ u_{m,n},v_{n,m}\right\} =\left\{ A,B\right\} \exp {(-a|n|)}\exp {%
(-a|m|)},  \label{eq5}
\end{equation}%
where $A$ and $B$ (but not $a$, see below) are treated as variational
parameters. This form of the trial function admits different amplitudes, $%
A\neq B$, of the solutions in the coupled lattices, but postulates equal
widths in both of them, $a^{-1}$. In this context, the SSB is signaled by
the emergence of asymmetric solutions, with $A^{2}\neq B^{2}$ \cite%
{VA,reshetke}.

The inverse width $a$ of the localized trial solution is found independently
from the linearization of Eqs. (\ref{eq2}), which is valid in the soliton's
tails (at $|m|,|n|\rightarrow \infty $),
\begin{equation}
a=-\ln \left( (4-\mu )/4-\sqrt{(4-\mu )^{2}/16-1}\right) ,  \label{eq6}
\end{equation}%
provided that the propagation constant is negative, $\mu <0$ (otherwise, the
solution cannot be localized). Relation (\ref{eq6}) may also be cast in
another form, that will be used below:
\begin{equation}
s\equiv e^{-a}=\frac{4-\mu }{4}-\sqrt{\frac{(4-\mu )^{2}}{16}-1},~\mu
=4-2(s+s^{-1}).  \label{eq7}
\end{equation}

The substitution of ansatz (\ref{eq5}) into Eqs. (\ref{eq3}) and (\ref{eq4})
yields the corresponding effective Lagrangian with two variational
parameters $A$ and $B$, where Eq. (\ref{eq7}) is used to eliminate $\mu $ in
favor of $s$:
\begin{equation}
L_{\mathrm{eff}}=\left( L_{u}\right) _{\mathrm{eff}}+\left( L_{v}\right) _{%
\mathrm{eff}}+2\varepsilon AB,  \label{eq8}
\end{equation}%
\begin{eqnarray}
\left( L_{u}\right) _{\mathrm{eff}} &=&-2A^{2}\frac{1+s^{2}}{s}+\frac{1}{2}%
A^{4}\frac{(1+s^{4})^{2}}{(1-s^{4})^{2}},  \nonumber \\
\left( L_{v}\right) _{\mathrm{eff}} &=&-2B^{2}\frac{1+s^{2}}{s}+\frac{1}{2}%
B^{4}\frac{(1+s^{4})^{2}}{(1-s^{4})^{2}}.  \label{eq9}
\end{eqnarray}%
The Euler-Lagrange equations for amplitudes $A$ and $B$ are $\left( \partial
/\partial A\right) \left( L_{u}\right) _{\mathrm{eff}}+2\varepsilon
B=0,\left( \partial /\partial B\right) \left( L_{v}\right) _{\mathrm{eff}%
}+2\varepsilon A=0$, or, in the explicit form,%
\begin{eqnarray}
-2\frac{1+s^{2}}{s}A+\frac{(1+s^{4})^{2}}{(1-s^{4})^{2}}A^{3}+\varepsilon B
&=&0  \nonumber \\
-2\frac{1+s^{2}}{s}B+\frac{(1+s^{4})^{2}}{(1-s^{4})^{2}}B^{3}+\varepsilon A
&=&0.  \label{eq10}
\end{eqnarray}%
These equations allow us to predict the existence of three different types
of the complexes formed by the fundamental localized modes centered at the
linkage site: symmetric and antisymmetric ones, with $A=B$ and $A=-B$,
respectively, and asymmetric modes with $A^{2}\neq B^{2}$.

\subsubsection{Existence regions for the interface soliton complexes}

\begin{figure}[ht]
\center\includegraphics [width=13cm]{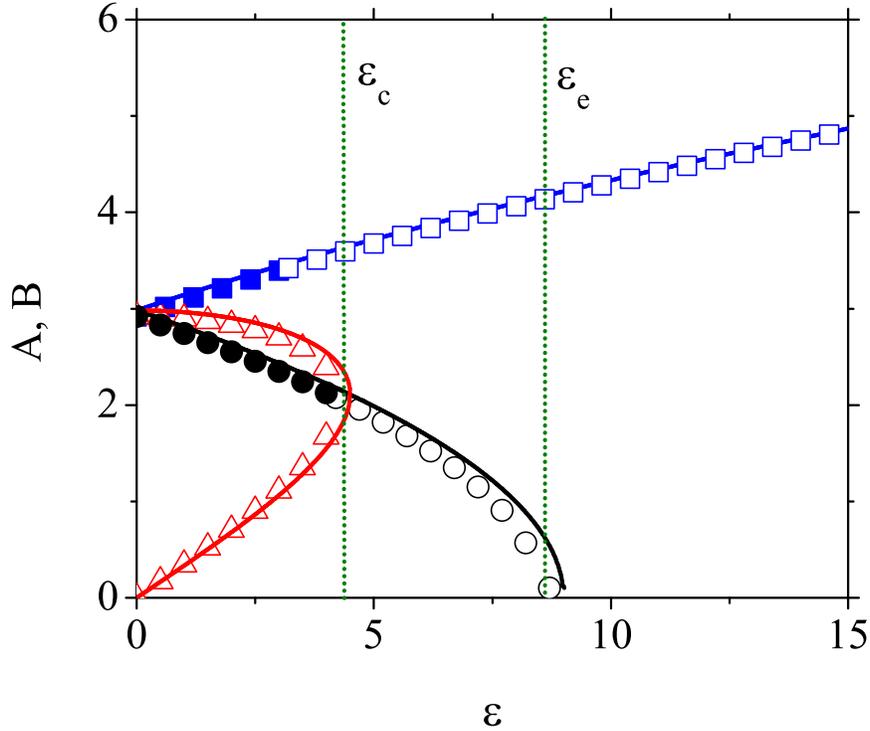}
\caption{(Color online) Amplitudes $A$ and $B$ of asymmetric, symmetric, and
antisymmetric solitons (red, black, and blue colors, respectively), as
predicted by the variational approximation (lines) and obtained in the
numerical form (triangles, circles, and squares pertain to the asymmetric,
symmetric, and antisymmetric modes, respectively) vs. the inter-lattice
linkage strength $\protect\varepsilon $. The propagation constant is fixed
to $\protect\mu =-5$. The dotted green vertical lines denote the numerically
found critical values of $\protect\varepsilon $ limiting the existence
regions of the asymmetric ($\protect\varepsilon _{c}$) and symmetric ($%
\protect\varepsilon _{e}$) solitons. Filled and empty symbols correspond to
unstable and stable solitons, respectively, as concluded from the numerical
investigation.}
\label{fig2}
\end{figure}

\begin{figure}[ht]
\center\includegraphics [width=12cm]{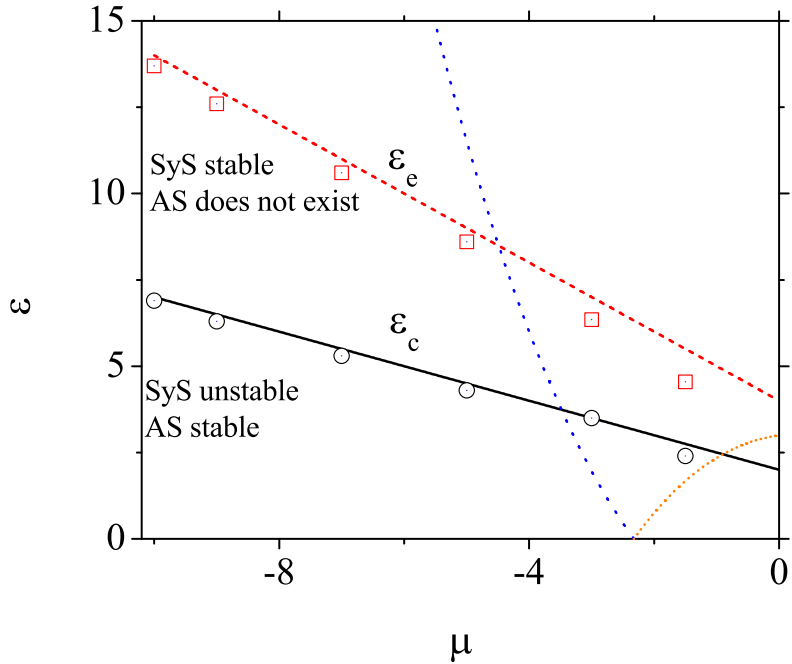}
\caption{(Color online) The existence and stability diagrams for the
fundamental symmetric (SyS), asymmetric (AS) and antisymmetric (AnS)
solitons. The variational results are presented by curves, and numerical
findings---by symbols in the parameter space $\left( \protect\varepsilon ,%
\protect\mu \right) $. Black circles and the corresponding line mark the
boundary of the AS existence region. Red squares and the dashed line denote
the existence boundary for the SyS modes. The numerical calculations show
that the symmetry-breaking bifurcation takes place along the curve $\protect%
\varepsilon _{c}(\protect\mu )$. The AnS mode exists in the entire parameter
plane. According to the Vakhitov-Kolokolov criterion, the AS may be stable
in the whole existence region, the SyS may be stable above the orange dashed
line, and AnS---to the left from the blue dotted line. The stability type of
the SyS and AS modes, indicated in the figure, was established by means of
numerical computations.}
\label{fig3}
\end{figure}

The solution for the symmetric soliton complexes is easily obtained from
Eqs. (\ref{eq10}):
\begin{equation}
A^{2}=\frac{(1-s^{4})^{2}}{(1+s^{4})^{2}}\left[ \frac{2}{s}%
(1+s^{2})-\varepsilon \right] .  \label{eq11}
\end{equation}%
As follows from Eq. (\ref{eq11}), the existence domain of the symmetric
solutions is $\varepsilon <${$\varepsilon _{e}\equiv 2(1+s^{2})/s$. For $\mu
=-5$, solution branches of all the types, produced by the VA along with
their numerical counterparts, are shown in Fig. \ref{fig2}, and the
respective existence regions, including the one given by curve $\varepsilon
_{e}(\mu )$, is displayed in Fig. \ref{fig3}. The procedure for obtaining
numerical results is described below. }

The amplitudes and existence range of the asymmetric solution complexes,
with $A^{2}\neq B^{2}$, can be calculated by adding and subtracting Eqs. 
(\ref{eq10}). After a straightforward procedure, the following expressions
for the amplitudes are obtained:
\begin{eqnarray}
A &=&\pm \frac{(1-s^{4})}{(1+s^{4})\sqrt{s}}\sqrt{1+s^{2}+\sqrt{%
(1+s^{2})^{2}-\varepsilon ^{2}s^{2}}},  \nonumber \\
B &=&\frac{(1-s^{4})^{2}}{(1+s^{4})^{2}}\frac{\varepsilon }{A}.  \label{eq12}
\end{eqnarray}%
This soliton mode exists at ${\varepsilon <\varepsilon }_{c}\equiv
(1+s^{2})/s$, where ${\varepsilon }_{c}$ is the bifurcation value. The
existence of the asymmetric mode may be naturally expected when the linkage
constant ($\varepsilon $) is not too large \cite{VA}, \cite{Akhmediev}-\cite%
{2D-BEC}. Indeed, the ultimate asymmetric solution, with $A\neq 0$ and $B=0$%
, is obviously possible in the limit of $\varepsilon =0$, which corresponds
for the decoupled lattices. It is also natural that, with the decrease of $%
\varepsilon $, the symmetry-breaking bifurcation should occur at some $%
\varepsilon =\varepsilon _{c}$, where the two asymmetric branches emerge
from the symmetric one, see Fig. \ref{fig2}. However, the stability of the
related solutions cannot be predicted solely by the VA.

For antisymmetric soliton complexes, the amplitude is obtained from Eq. (\ref%
{eq10}) by setting $A=-B$, which yields
\begin{equation}
A^{2}=\frac{(1-s^{4})^{2}}{(1+s^{4})^{2}}\left[ \frac{2}{s}%
(1+s^{2})+\varepsilon \right] ,  \label{eq13}
\end{equation}%
cf. Eq. (\ref{eq11}) for the symmetric modes. Relation (\ref{eq13}) is
plotted versus $\varepsilon $ for fixed $\mu =-5$ by the blue (upper) curve
in Fig. \ref{fig2}. As follows from this relation, the VA predicts the
existence of the antisymmetric solitons in the entire parameter space, on
the contrary to the limited existence regions predicted for the symmetric
and asymmetric modes.

\subsubsection{Stability of the soliton complexes}

The stability of the discrete solitons predicted by the VA can be estimated
by dint of the VK criterion, $dP/d\mu >0$, where $P\equiv \sum_{n=-\infty
}^{n=+\infty }\sum_{m=-\infty }^{m=+\infty }(u_{n,m}^{2}+v_{n,m}^{2})$ is
the total norm (power) of the soliton complex \cite{isrl}. The norm
corresponding to the ansatz (\ref{eq5}) is
\begin{equation}
P=(A^{2}+B^{2})\left( \frac{1+s^{2}}{1-s^{2}}\right) ^{2}.  \label{eq14}
\end{equation}%
For the symmetric solution with $A=B,$ Eqs. (\ref{eq11}) and (\ref{eq14})
yield
\begin{equation}
P=2\frac{(1+s^{2})^{4}}{(1+s^{4})^{2}}\left[ \frac{2}{s}(1+s^{2})-%
\varepsilon \right] .  \label{eq15}
\end{equation}%
This expression satisfies condition $\partial P/\partial s<0$, which is
tantamount to the VK criterion, in the region of
\begin{equation}
\varepsilon >\frac{2(1+s^{2})}{s}-\frac{(1+s^{2})(1+s^{4})}{4s^{3}}.
\label{eq16}
\end{equation}%
This region is displayed in Fig. (\ref{fig3}) as the area above the dashed
orange curve (in the right bottom corner of the figure). Thus, according to
the VK criterion, the stable symmetric branch may exist in the large part of
the parameter plane, except for the small region below the dashed orange
curve, which corresponds to the weak inter-lattice linkage and wide solitons.

For the asymmetric solutions, the substitution of Eq. (\ref{eq12}) into Eq. (%
\ref{eq14}) produces a simple expression for the total norm, which does not
depend on $\varepsilon $:
\begin{equation}
P=2\frac{(1+s^{2})^{5}}{s(1+s^{4})^{2}}.  \label{eq17}
\end{equation}%
It also satisfies the VK criterion in the entire region of the existence of
the asymmetric mode.

For the antisymmetric modes with $A=-B$, the use of Eq. (\ref{eq13}) gives
\begin{equation}
P=2\frac{(1+s^{2})^{4}}{(1+s^{4})^{2}}\left[ \frac{2}{s}(1+s^{2})+%
\varepsilon \right] .  \label{eq18}
\end{equation}%
In this case, condition $dP/ds<0$ is satisfied at
\begin{equation}
\varepsilon <\frac{(1+s^{2})(1+s^{4})}{4s^{3}}-\frac{2(1+s^{2})}{s}.
\label{eq19}
\end{equation}%
The region defined by Eq. (\ref{eq19}) is shown in Fig. (\ref{fig3}) as the
area to the left from the blue dotted curve (the one which cuts the entire
plane).

However, the VK criterion offers only the necessary condition for the
soliton stability. To identify regions of full stability of the soliton
complexes, the VK criterion should be combined with the spectral condition,
which requires the existence of only pure imaginary eigenvalues in the
linearization of Eqs. (\ref{eq1}) with respect to small perturbations around
the stationary soliton solutions. The spectral analysis is performed
numerically in the next section. In fact, the results demonstrate that,
unlike the prediction of the shape of the soliton modes by means of the VA,
the stability prediction based on the formal application of the VA criterion
is not accurate.

\section{Numerical results}

The predictions of the VA were verified by numerically solving stationary
equations (\ref{eq2}). The numerical algorithm is based on the modified
Powell minimization method \cite{nashi2d}. The initial guess to construct
soliton complexes centered at the linkage site of the 2D lattices 
(Fig. \ref{fig1}) was taken as $u_{0}=v_{0}=A>0$ for symmetric solutions, $%
u_{0}=A>0,\,v_{0}=B\neq A>0$ for asymmetric ones, and $u_{0}=A>0,\,v_{0}=-A$
for solutions of the antisymmetric type, with the VA-predicted values of $A$
and $B$. At other sites, the amplitudes of the initial guess are set to be
zero.

\subsection{Stationary soliton modes}

The stability of the stationary modes was investigated through the
calculation of eigenvalues (EVs) of small perturbations around the
stationary solutions, which were computed following the lines of Refs. \cite%
{nashi,nashi2d,reshetke}. The obtained results were further verified in
direct numerical simulations of Eqs. (\ref{eq1}), using the sixth-order
Runge-Kutta algorithm, as in Refs. \cite{nashi2d,reshetke}. The simulations
were carried out for stationary soliton complexes to which initial
perturbations were added.

\begin{figure}[ht]
\center\includegraphics [width=6cm]{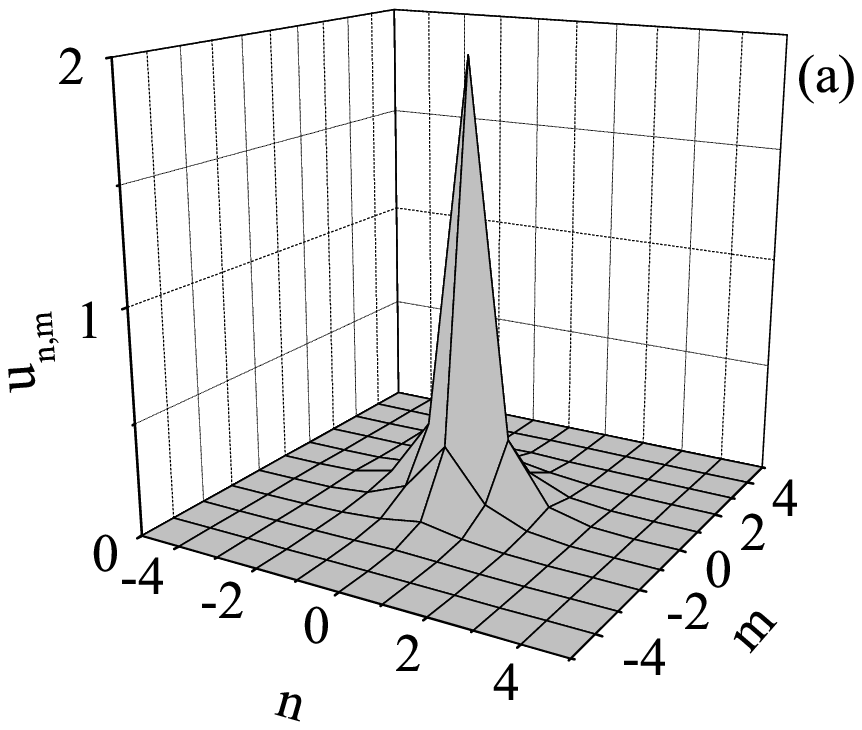}%
\includegraphics
[width=6cm]{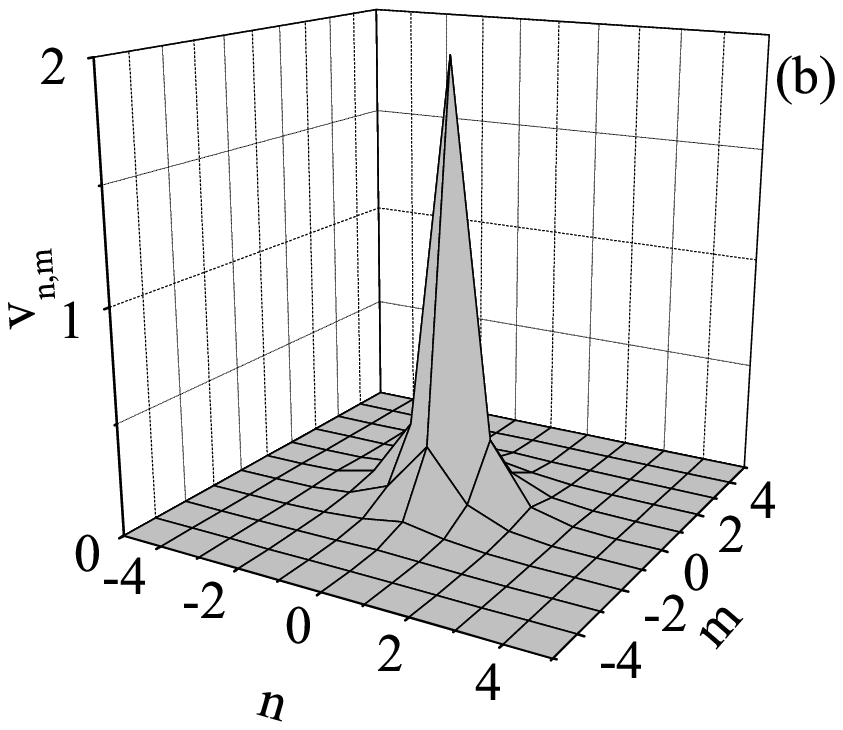}\newline
\includegraphics
[width=6cm]{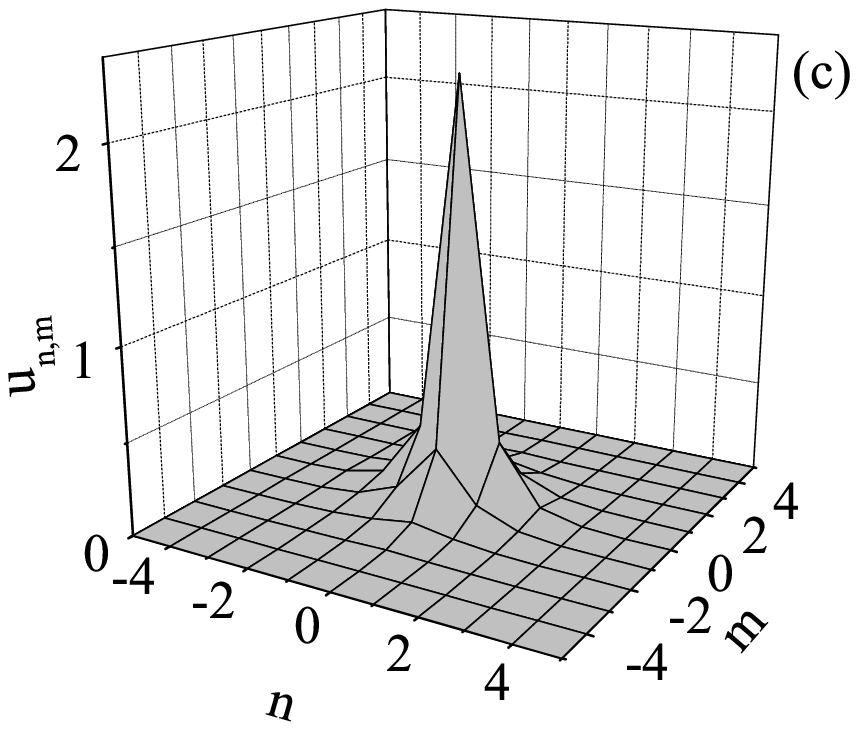}\includegraphics
[width=6cm]{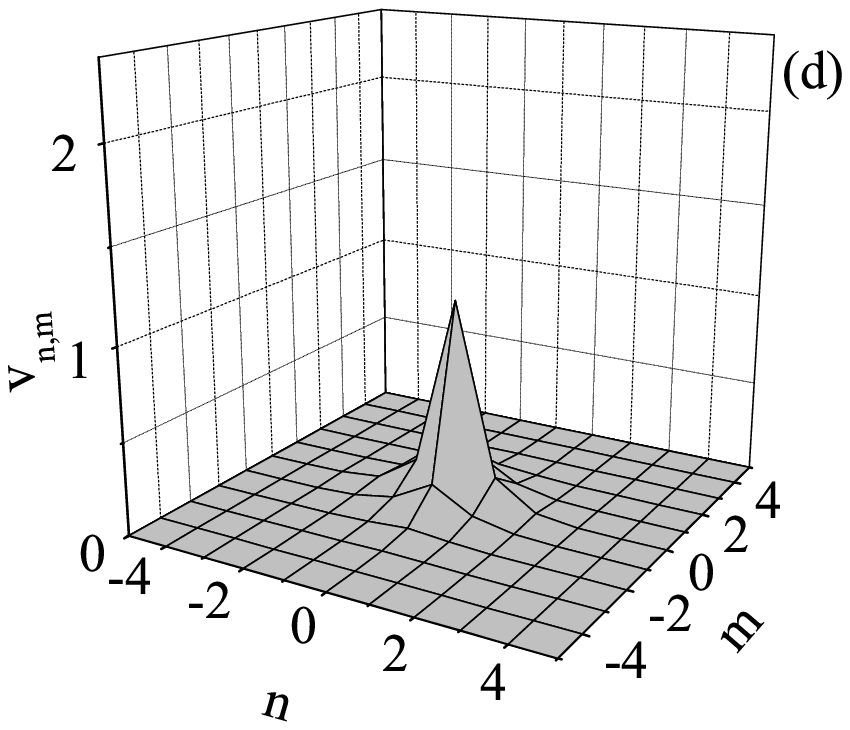}\newline
\includegraphics [width=6cm]{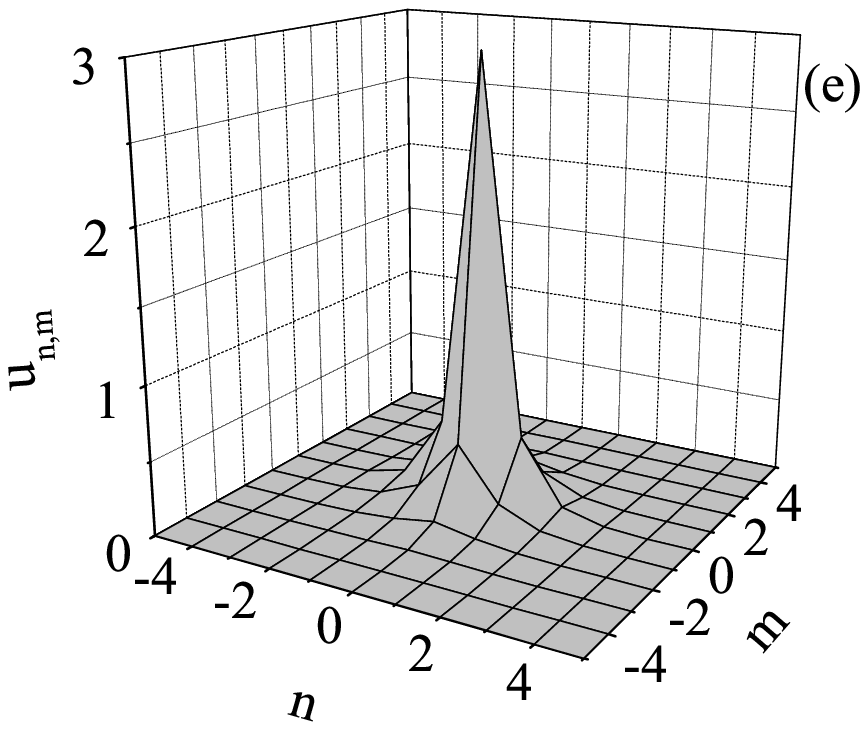}%
\includegraphics
[width=6cm]{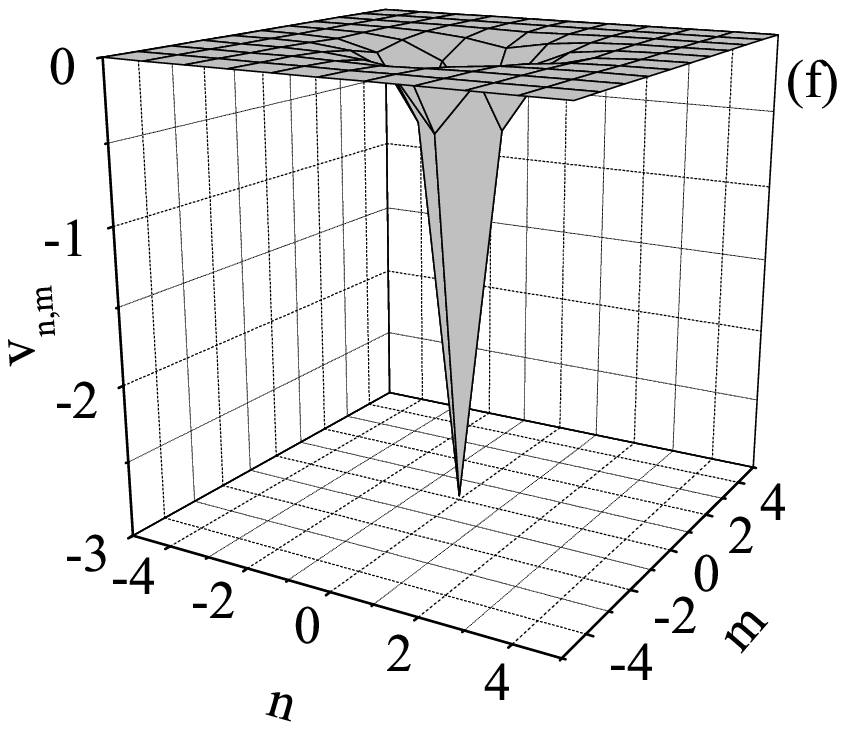}
\caption{Typical shapes of the soliton complexes of different types:
symmetric in (a) and (b), asymmetric in (c) and (d), and antisymmetric in
(e) and (f). The parameters are $\protect\mu =-3$, $\protect\varepsilon =2.5$%
, the corresponding norms for complexes being $P_{\mathrm{symm}}=8.63,\,P_{%
\mathrm{asymm}}=7.11$, and $P_{\mathrm{antisymm}}=19.91$. }
\label{fig4}
\end{figure}

Typical shapes of symmetric, asymmetric and antisymmetric soliton complexes
found in the numerical form are displayed in Fig. \ref{fig4}. The respective
dependencies of the amplitudes $A$ and $B$ on $\varepsilon $ were displayed
above, for all the soliton branches, alongside their VA-predicted
counterparts, in Figs. \ref{fig2} and \ref{fig3}. The numerical results show
that the symmetric and asymmetric complexes exist in bounded regions of the
parameter space, see Fig. \ref{fig3}. The results predicted by means of the
VA are in good agreement with their numerical counterparts for all the types
of the soliton complexes. The\ comparison of the variational and numerical
dependencies of the solitons' amplitudes on $\varepsilon $, for fixed $%
s\approx 0.23$, which corresponds to $\mu =-5$, can be seen in Fig. \ref%
{fig2}. The difference between the analytical and numerical results is
negligible for small $\varepsilon $, and slightly grows with $\varepsilon $.
In particular, the existence border of the symmetric complexes predicted by
the VA is $\varepsilon _{e}\equiv 9$, while its numerical counterpart is $%
\varepsilon _{e}\equiv 8.63$.

Finally, it is easy to check that both the variational and numerical results
demonstrate that the SSB pitchfork bifurcation observed in Fig. \ref{fig2}
is of the \textit{supercritical} type, similarly to that reported in Ref.
\cite{reshetke} for the 1D counterpart of the present system (in contrast
with the \textit{subcritical} bifurcation demonstrated by the system of 1D
and 2D parallel chains with the uniform linear coupling acting at each site
\cite{VA}).

\subsection{The stability analysis}

\begin{figure}[ht]
\center\includegraphics
[width=6cm]{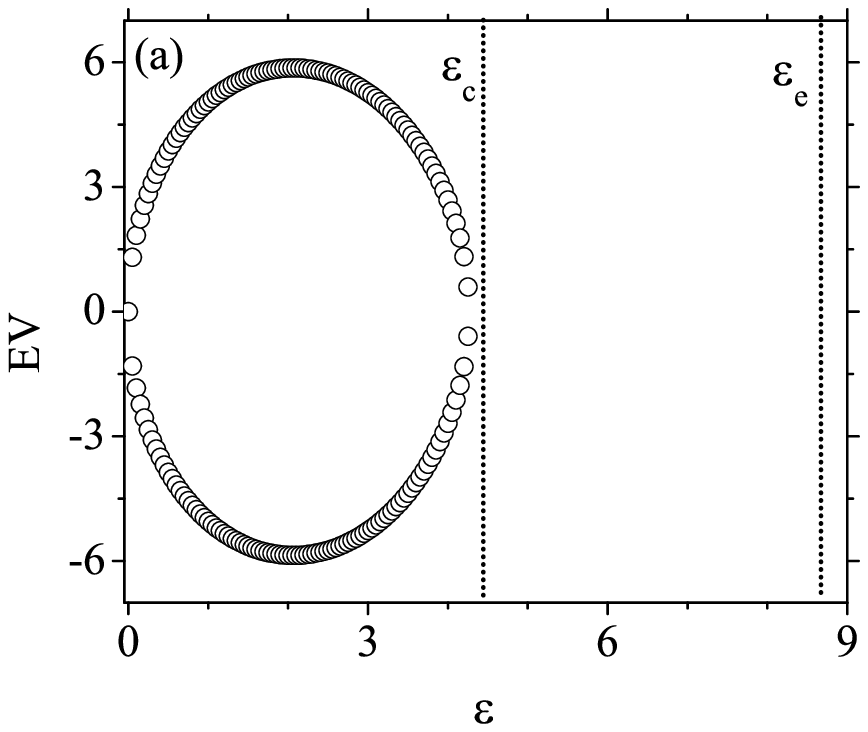}\includegraphics
[width=6cm]{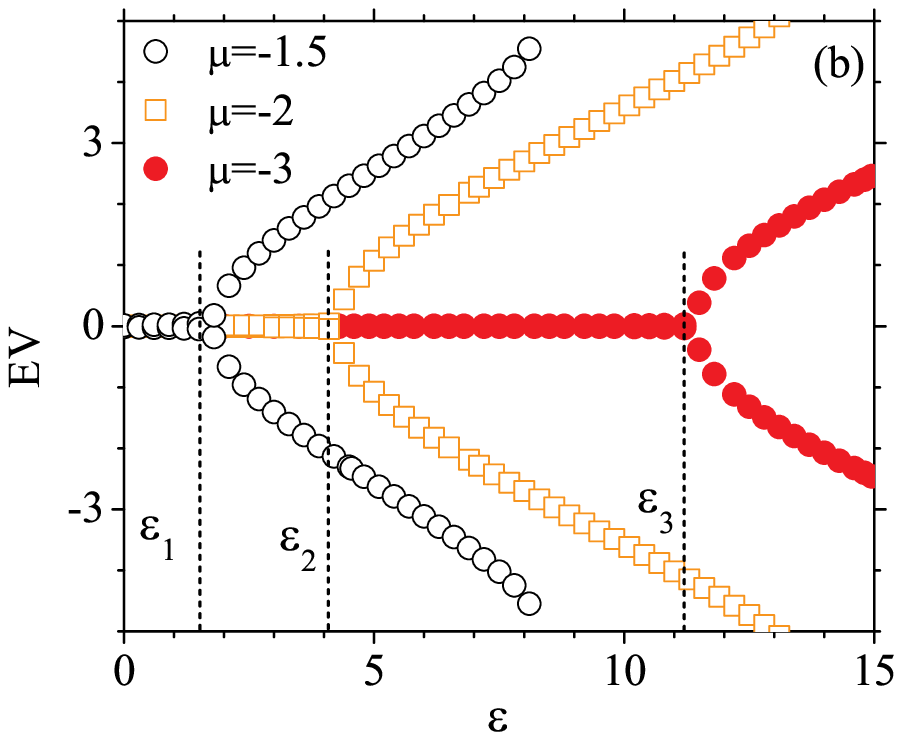}\includegraphics
[width=6cm]{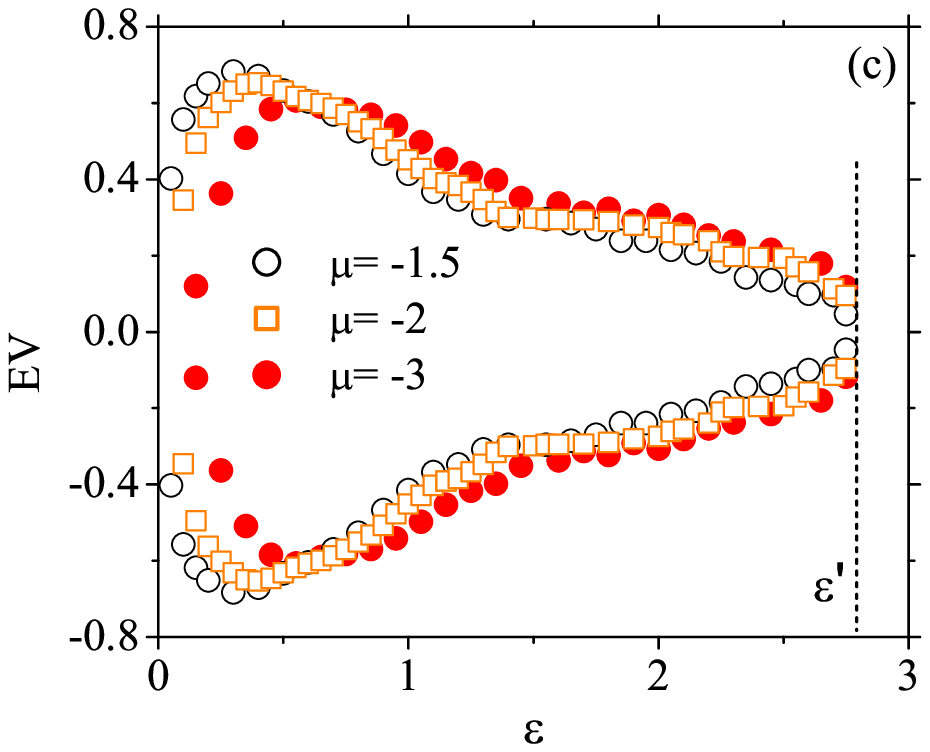}
\caption{(Color online) (a) Pure real eigenvalues (EVs ) vs. $\protect%
\varepsilon $ for symmetric solitons with $\protect\mu =-5$. Dotted lines
denote the bifurcation value $\protect\varepsilon _{c}$ and the existence
threshold $\protect\varepsilon _{e}$ for the symmetric solitons. Pure real
EVs, and the real part of complex EVs, are shown vs. $\protect\varepsilon $
in panels (b) and (c), respectively, for antisymmetric solitons. Black (empty circles),
orange (squares), and red (filled circles) symbols correspond, respectively, to fixed $\protect\mu %
=-1.5,-2,$ and $-3$. Dotted lines marked by $\protect\varepsilon _{1,2,3}$
in plot (b), and by $\protect\varepsilon ^{\prime }$ in plot (c) are
boundaries of regions where the pure real EVs (b), or real parts of the
complex EVs (c), take significant values, $\mathrm{Re}(\mathrm{EV})>0.001$.}
\label{fig5}
\end{figure}

The linear-stability analysis demonstrates that the symmetric complexes
emerge as stable solutions at $\varepsilon =\varepsilon _{e}$, and, as
expected, they change the stability at the bifurcation point, $\varepsilon
=\varepsilon _{c}$, where the asymmetric solution branches appear (see Figs. %
\ref{fig2} and \ref{fig3}). Unstable symmetric solutions are characterized
by the pure real EV pairs, see Fig. \ref{fig5}(a). For the parameter set
used in Fig. \ref{fig5}(a), the isolated discrete solitons existing in the
uncoupled lattices at $\varepsilon =0$ are stable \cite{pgk2d,nashi2d}. The
introduction of the linkage between the lattices changes the stability of
the symmetric complex formed by such solitons. The results \emph{do not}
confirm the prediction, based on the VK criterion, that the symmetric
complexes change the stability at the value of the $\varepsilon $ given by
Eq. (\ref{eq16}), which corresponds to the orange dashed curve in Fig. \ref{fig3}. 
The instantaneous destabilization of the symmetric bound states with
the increase of $\varepsilon $ from zero [see Fig. \ref{fig7}(a)] is simply
explained by the fact that symmetric complexes are always unstable against
the SSB in dual-core systems with a small linear-coupling constant \cite%
{Progress}.

\begin{figure}[h]
\center\includegraphics [width=6cm]{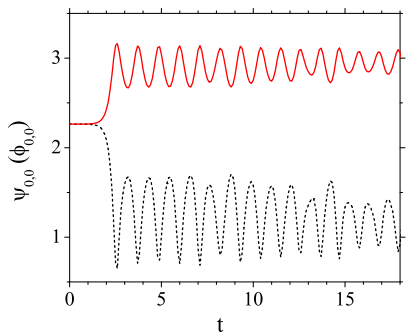}
\caption{The evolution of the amplitudes of the unstable SyS mode with $%
\protect\mu =-5$, $\protect\varepsilon =3.4$ into the corresponding stable
AS complex. The amplitudes of the components of the latter complex are
represented by different lines.}
\label{fig6}
\end{figure}

\begin{figure}[ht]
\center\includegraphics [width=4.5cm]{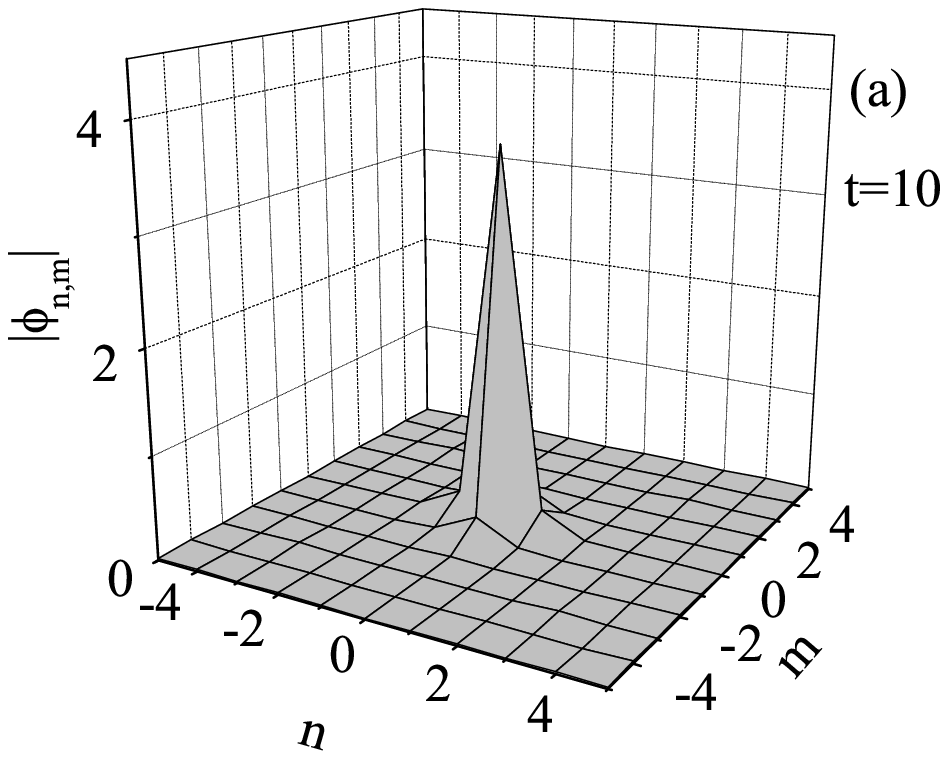}%
\includegraphics
[width=4.5cm]{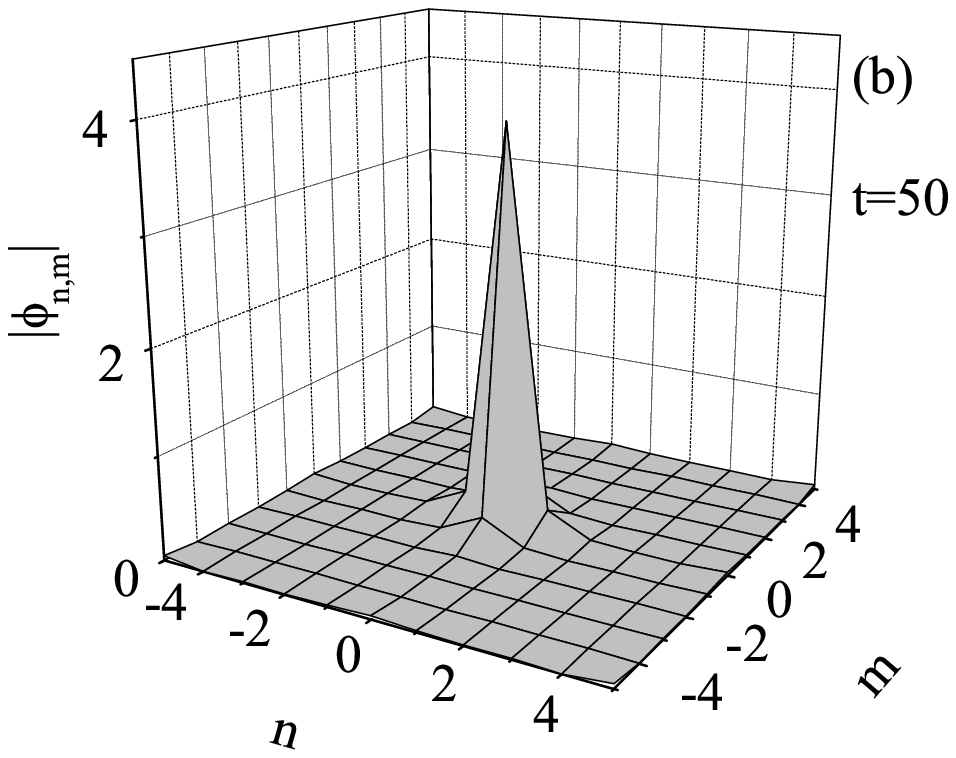}\includegraphics
[width=4.5cm]{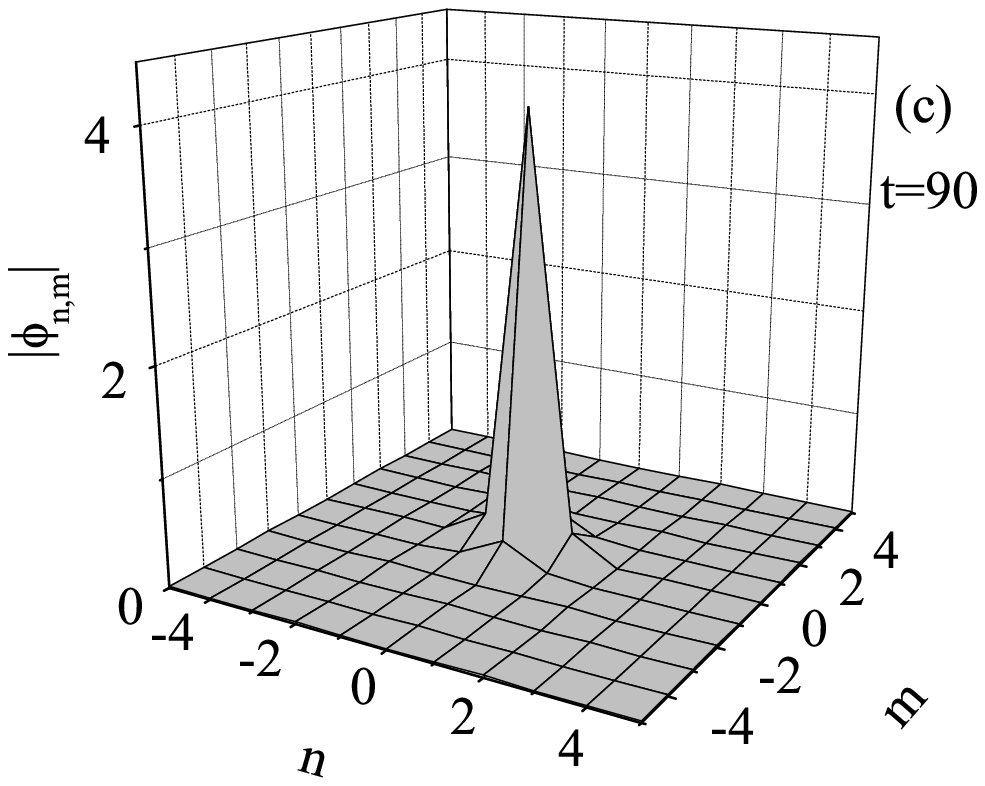}\includegraphics
[width=4.5cm]{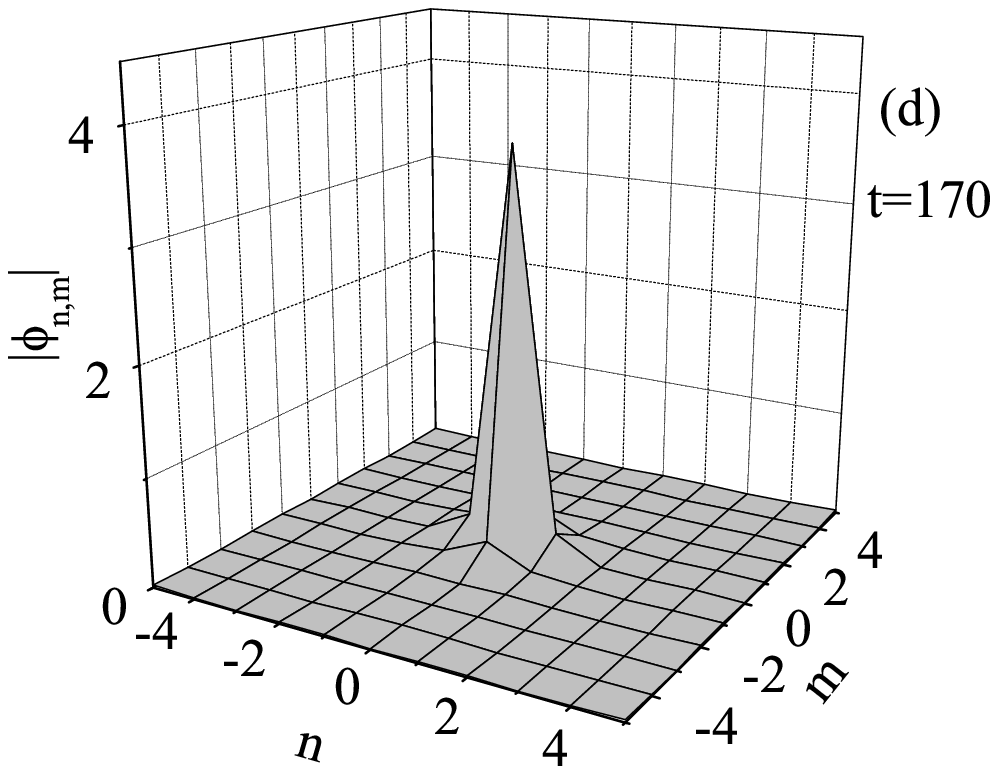}\newline
\includegraphics
[width=4.5cm]{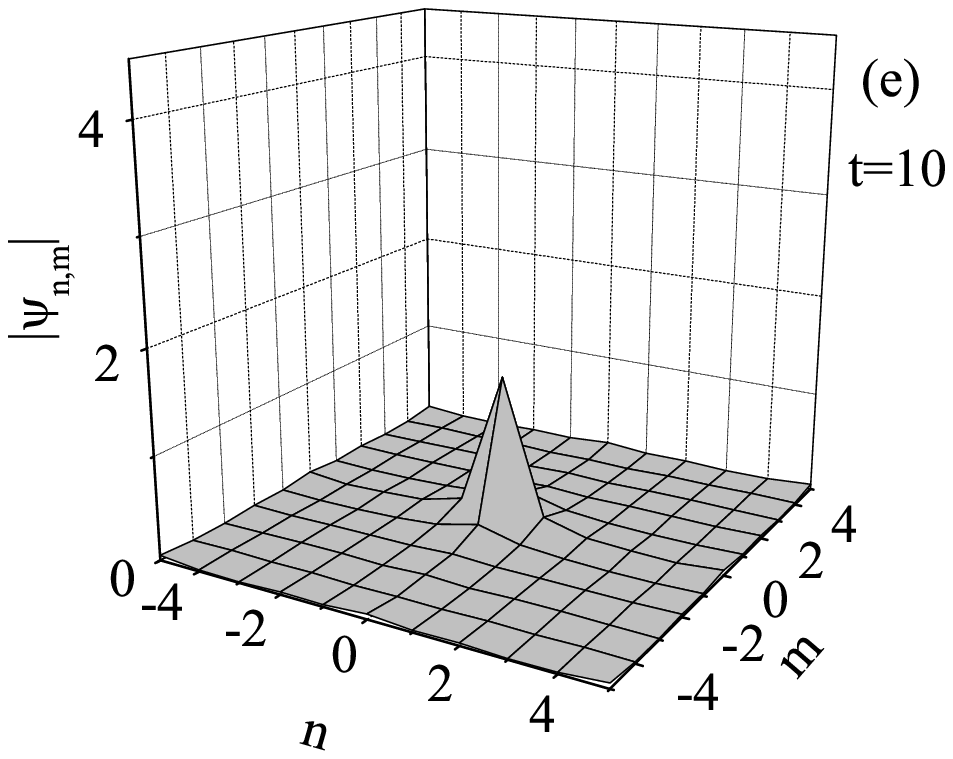}\includegraphics
[width=4.5cm]{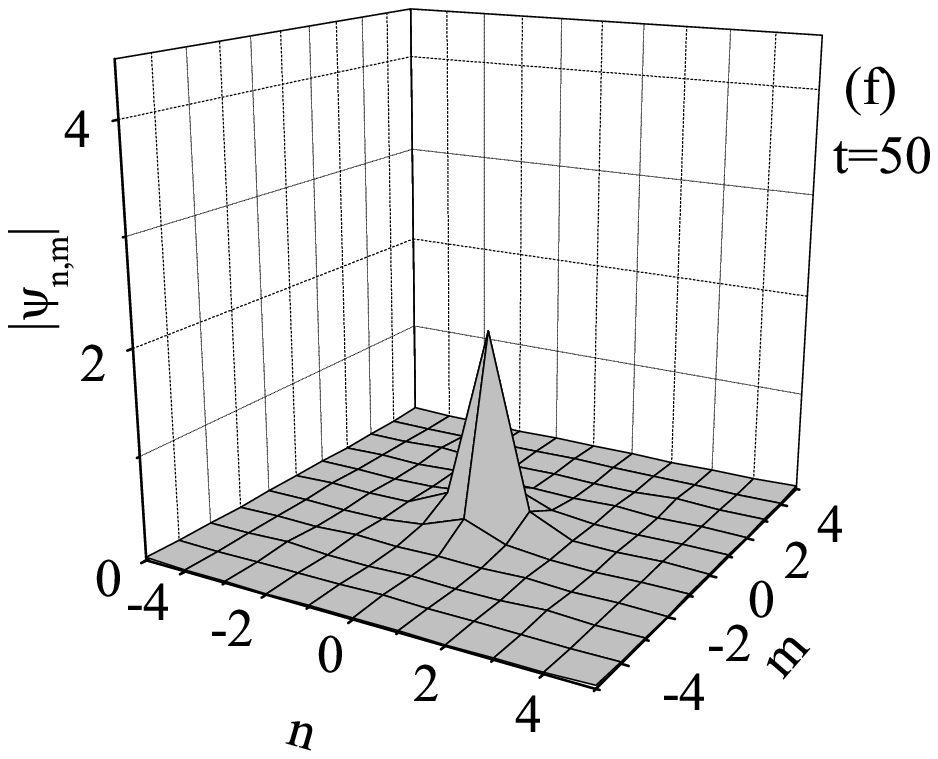}\includegraphics
[width=4.5cm]{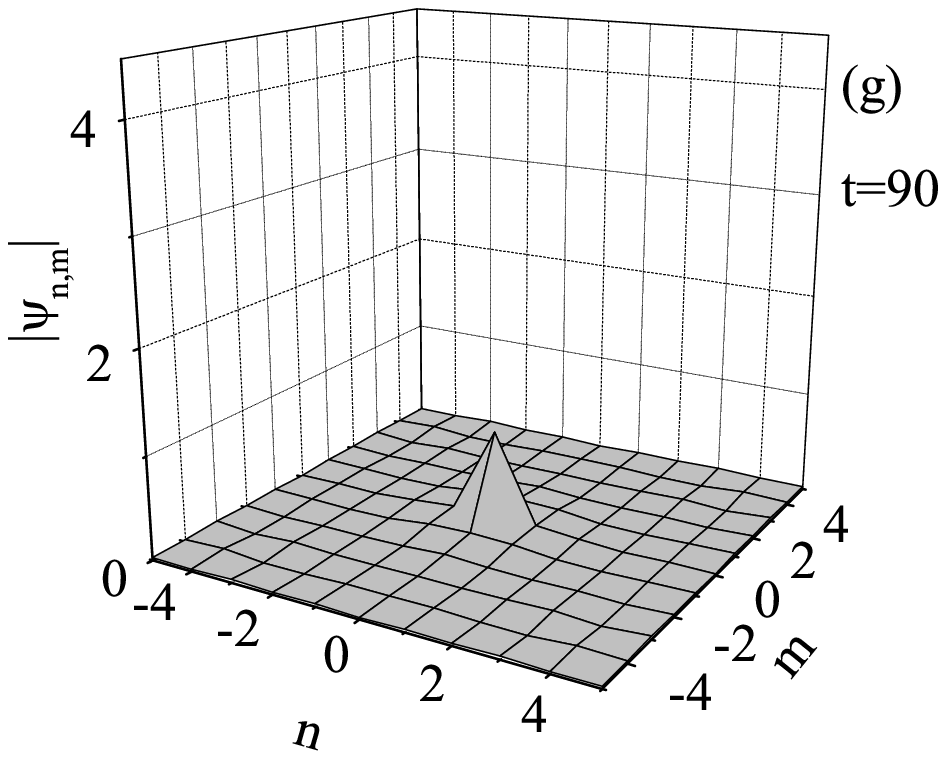}\includegraphics
[width=4.5cm]{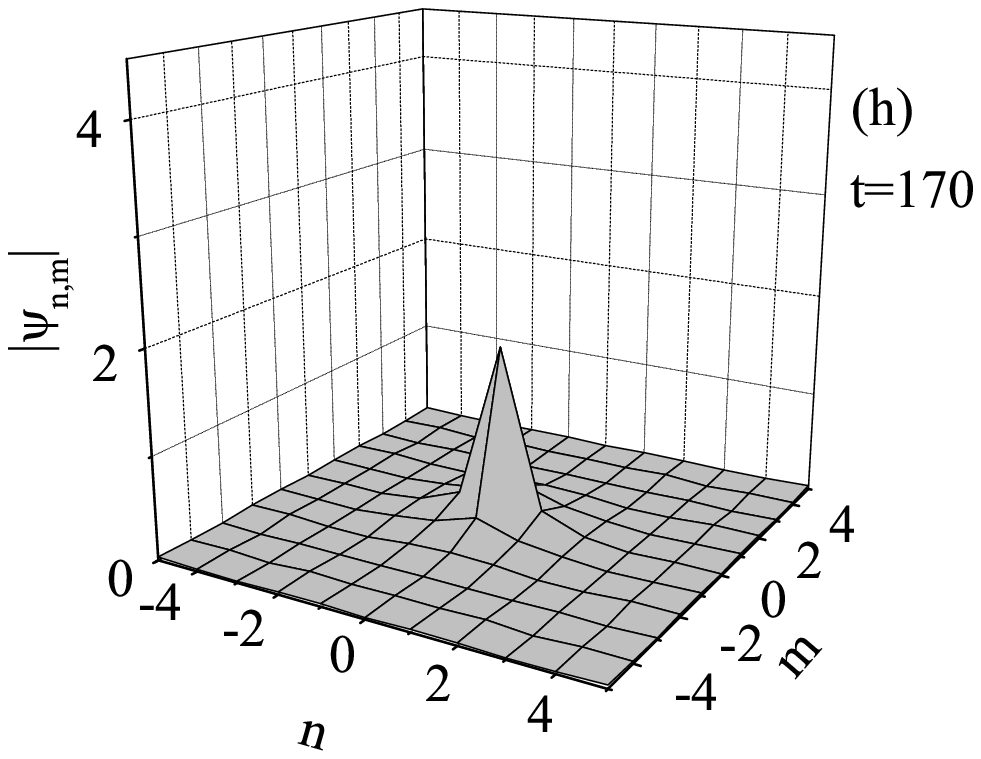}
\caption{The oscillatory evolution of a perturbed asymmetric complex with $%
P=12.02$, $\protect\varepsilon =5$, and $\protect\mu =-8$. The relative
strength of small perturbation with respect to the solution amplitude is $%
0.01$. Profiles of the two components are shown, in the top and bottom rows.}
\label{fig7}
\end{figure}

Direct simulations confirm the stability of the SyS complexes in the
interval of $\varepsilon _{c}<\varepsilon <\varepsilon _{e}$. On the other
hand, simulations of the evolution of unstable symmetric modes demonstrate
that, under the action of small perturbations, these unstable modes (at $%
\varepsilon <\varepsilon _{c}$) evolve into asymmetric breathing complexes,
which consist of two oscillating localized components, that exchange energy
in the course of the evolution. This behavior is illustrated in Fig. \ref{fig6}, 
where the evolution of the unstable SyS mode into the AS complex
is shown by plotting the amplitudes of the component solitons versus time.

Two mutually symmetric branches of asymmetric solutions, which are created
by the destabilization of the symmetric branch, see Eq. (\ref{eq12}), turn
out to be \emph{stable}, according to the linear-stability analysis, which
yields for them EV spectra with the zero real part. Direct simulations
corroborate that slightly perturbed asymmetric complexes are robust modes,
see Fig. \ref{fig7}.

\begin{figure}[ht]
\center\includegraphics [width=4cm]{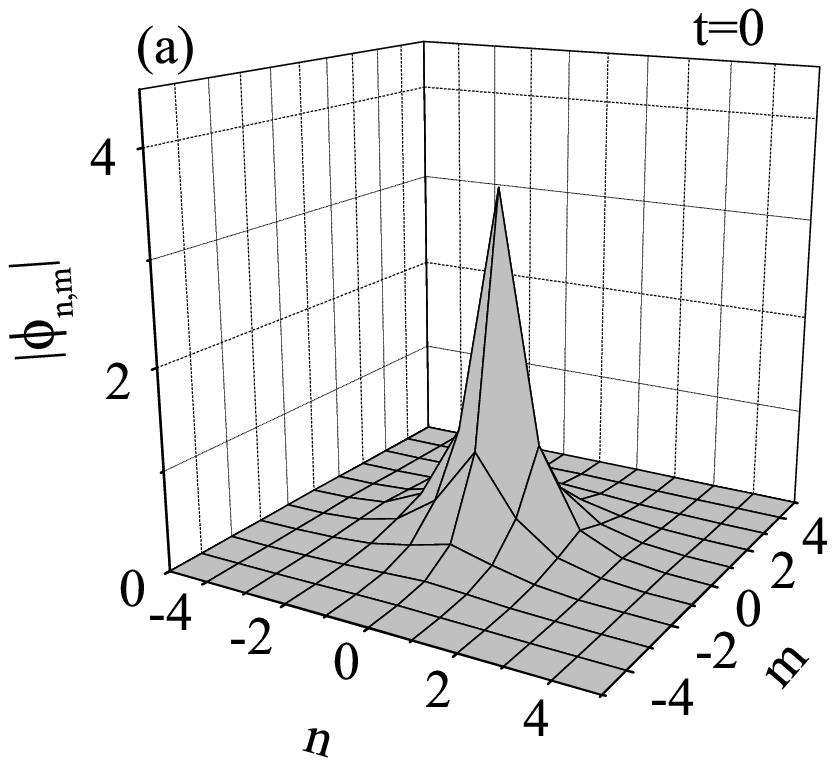}%
\includegraphics
[width=4cm]{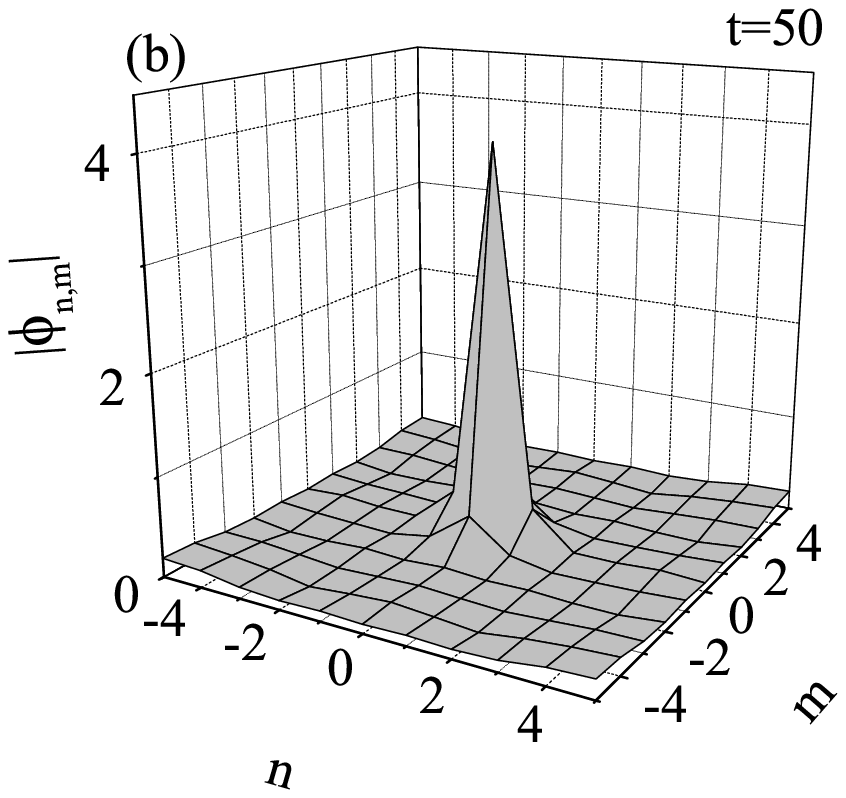} \includegraphics[width=4cm]{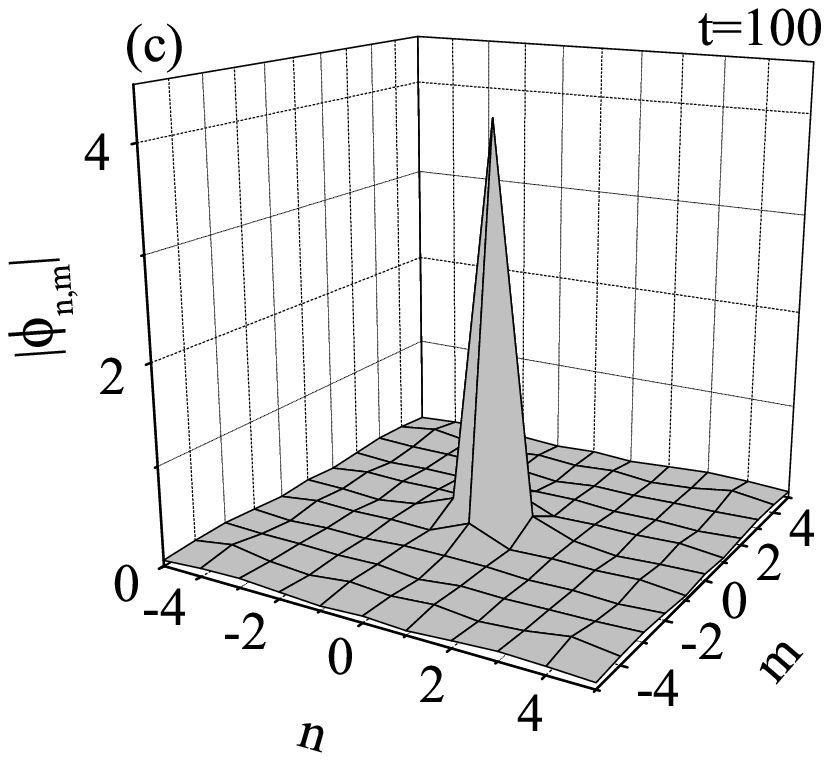} %
\includegraphics[width=4cm]{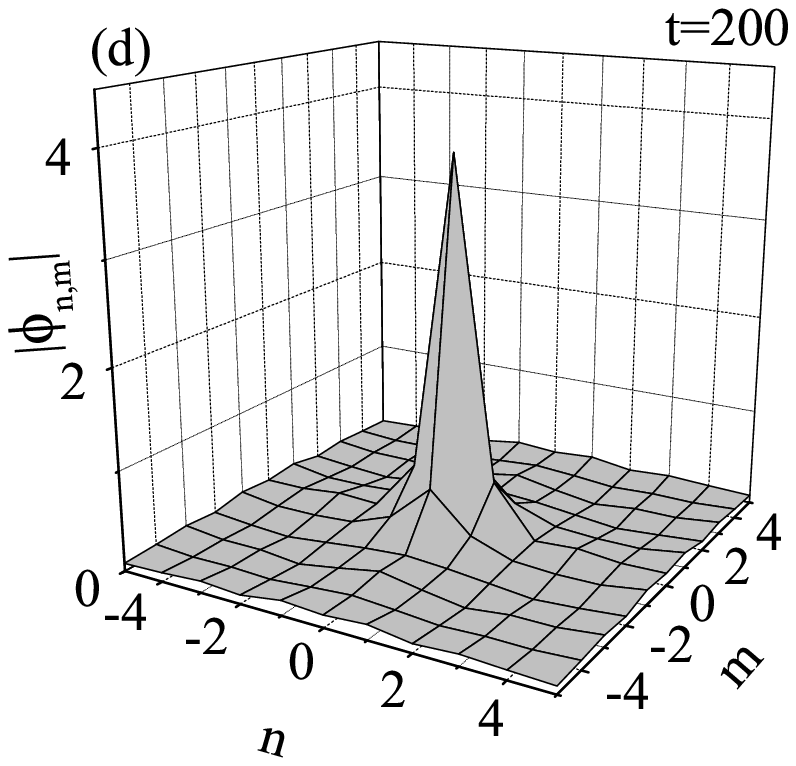}
\caption{The time snapshots illustrating the evolution of the AnS complex
with $\protect\mu =-1.5$, $\protect\varepsilon =7.2$, and $P=33.48$. Four
plots represent the amplitude profiles of the component solitons, $|\protect%
\phi _{m,n}|=|\protect\psi _{m,n}|$, at $t=0,50,100,$ and $200$. The
components are strongly pinned, with the amplitudes at the interface sites
slightly oscillating around value $3.8$. }
\label{fig8}
\end{figure}

The purely real EV pairs which are numerically calculated for the
antisymmetric modes become significant above some threshold values of $%
\varepsilon ,$ which depend on $\mu $. The threshold values, $\varepsilon
_{1,2,3}$ for $\mu =-1.5,-2,-3,$ are shown in Fig. \ref{fig5}(b). On the
other hand, the antisymmetric modes give rise to the complex EVs at $%
\varepsilon <\varepsilon ^{\prime }$ and arbitrary $\mu ,$ as shown in Fig. %
\ref{fig5}(c). Therefore, at large values of the coupling constant, $%
\varepsilon >\varepsilon _{1,2,3},$ the instability is determined by the
purely real EVs, but in the region of $\varepsilon <min(\varepsilon ^{\prime
},\varepsilon _{1,2,3})$ the real part of the complex EV dominates the
instability\textbf{.} Actually, the numerical results for the (in)stability
of the antisymmetric complexes do not corroborate the analytical prediction
presented by Eq. (\ref{eq19}).

The dynamics of the antisymmetric complexes with positive real parts of the
EVs is not significantly affected by small perturbations, except in the area
with very small $\varepsilon $, where the AnS, SyS and AS with close values
of the power coexist, see Fig. \ref{fig3} for $\varepsilon \ll \varepsilon
_{c}$. In the latter case, the AnS complexes follow the same scenario as the
unstable SyS complexes, i.e., the unstable AnS evolves into a breathing AS\
complex. The reason for the robustness of other antisymmetric modes is the
fact that the corresponding branch features large amplitudes of the
solitons, leading to their strong trapping at the central lattice sites.
Therefore, the actual instability (escape of the discrete wave fields) is
suppressed by the strong the Peierls-Nabarro potential barrier \cite{pn}.
Accordingly, due to the weak effective coupling between the central site and
the adjacent ones, the introduction of small perturbations can excite
internal oscillations but does not destroy the localization of the mode, and
the emerging breather (which keeps the antisymmetric structure, as concerns
the relation between its components) remains a strongly trapped state. This
is the case in \textbf{almost} the whole existence region of the
antisymmetric solitons. The dynamics of exponentially unstable antisymmetric
complexes (actually, with large pure real EVs) is illustrated in Fig. \ref{fig8} 
for $\mu =-1.5$ and $\varepsilon =7.2$, with norm $P=33.48$. Keeping
the antisymmetric structure, as said above, the two components feature
identical amplitude profiles, $|\psi _{n,m}|=|\phi _{n,m}|$. Both of them
shrink into more pinned modes that radiate away a small part of their norm
(energy), which forms a small but finite oscillating background, while the
central peak remains robust.

\begin{figure}[ht]
\center\includegraphics [width=4cm]{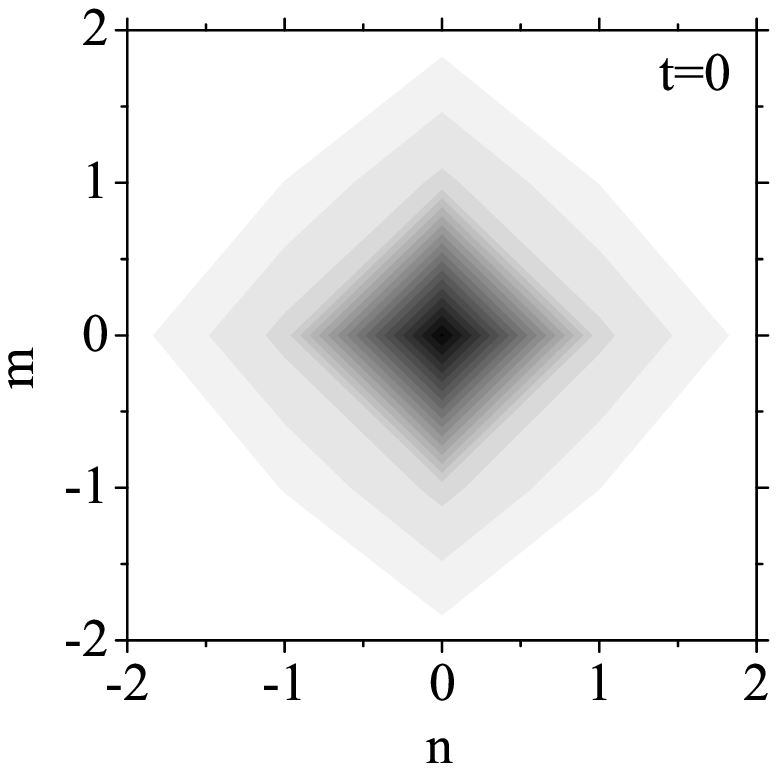}
\includegraphics
[width=4cm]{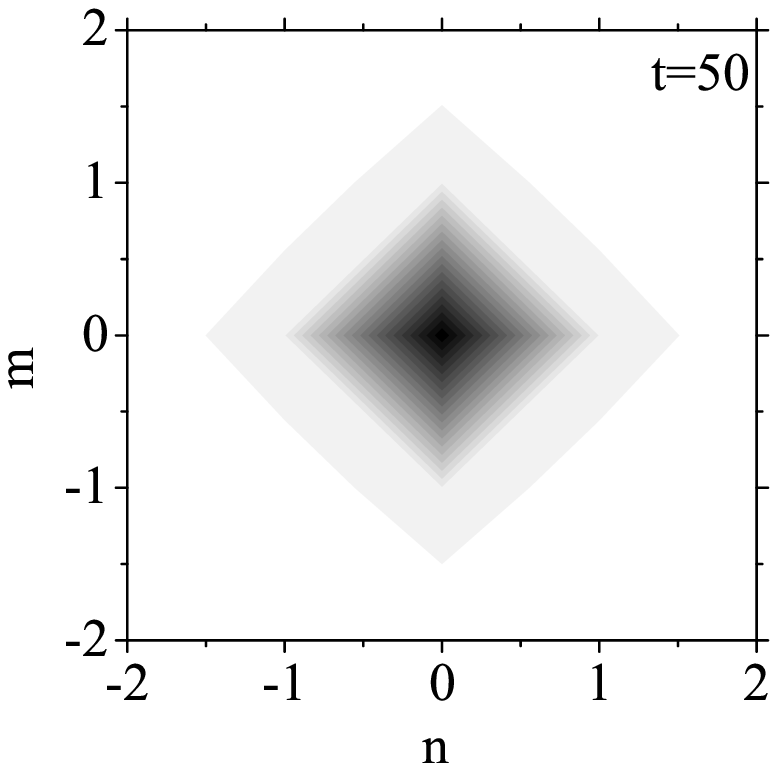}\includegraphics
[width=4cm]{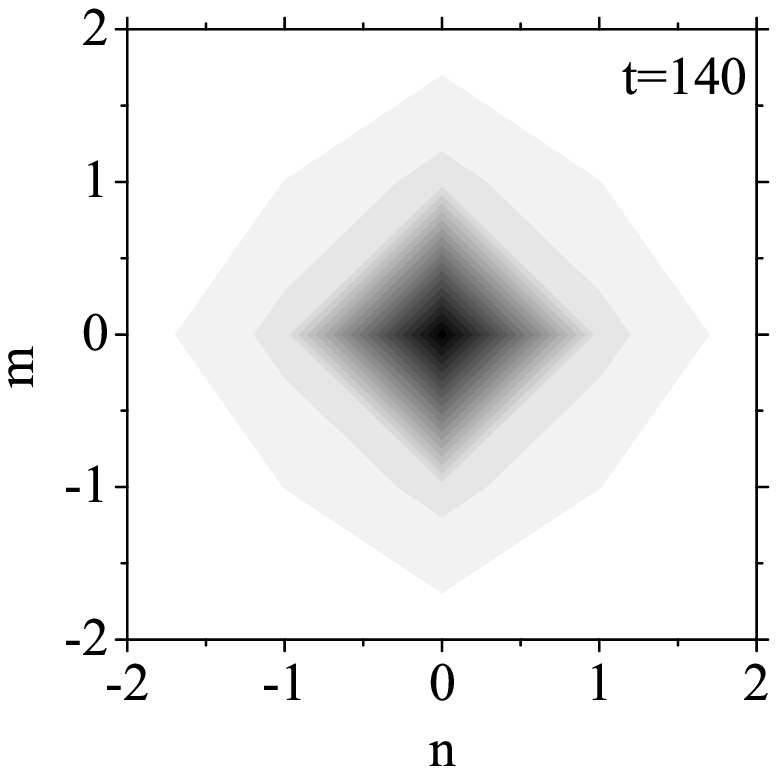}\includegraphics
[width=5cm]{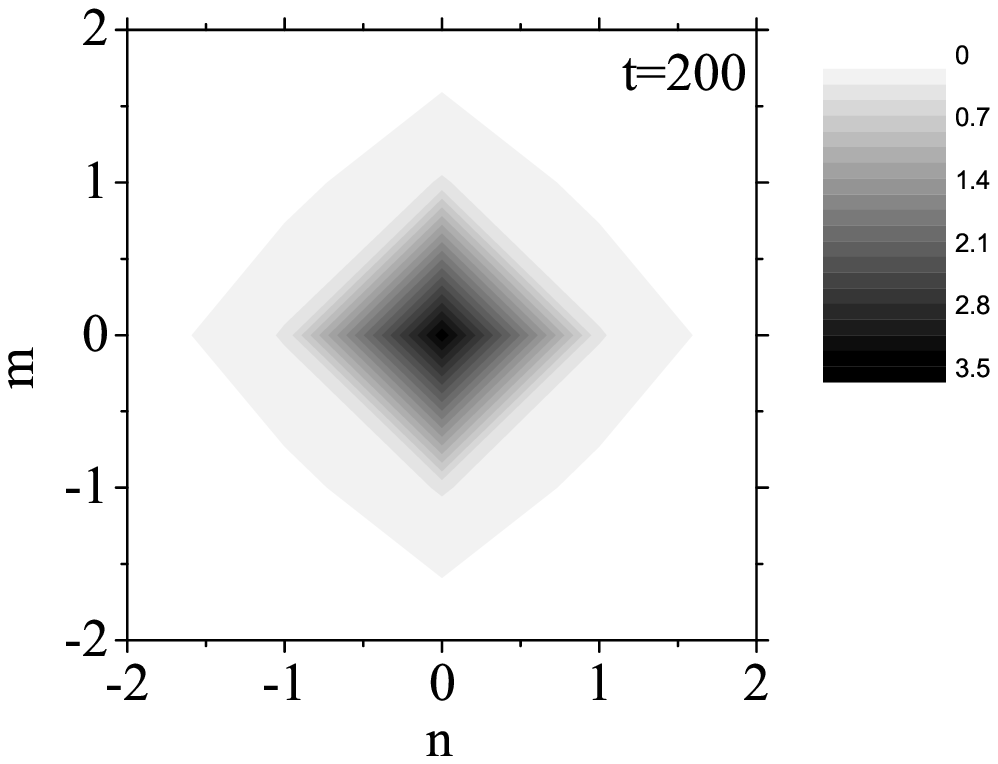}
\caption{The evolution of a perturbed antisymmetric mode which is unstable
against oscillatory perturbations, with $\protect\mu =-3$, $\protect%
\varepsilon =1.5$, and $P=17.64$. The component solitons are strongly
pinned, and their amplitudes slightly oscillate, similar to the case
displayed in Fig. \ref{fig8}. }
\label{fig9}
\end{figure}

The instability of the antisymmetric modes in the case of very small pure
real EVs develops very slowly, in the presence of small perturbations,
following the same scenario. In Fig. \ref{fig9}, the evolution of a typical
antisymmetric complex subject to the oscillatory unstable is shown. The part
of the initial energy lost into background is smaller then in the previous
case, and the central peak slightly oscillates in time.

Returning to the global existence diagrams, it is worth to note that two
bistability areas can be identified in them: the domain of the coexistence
of stable symmetric and quasi-stable antisymmetric solitons (the
quasi-stability pertains to very small growth rates mentioned above), or the
one featuring the simultaneous stability of asymmetric and antisymmetric
modes (at very small $\varepsilon $), on the opposite side on the SSB
bifurcation. This result is in accordance with similar findings reported in
other linearly-coupled two-component systems with the self-focusing
nonlinearity \cite{1D-BEC,2D-BEC,reshetke}.

It is also relevant to compare properties of the soliton complexes in the
present two-component 2D lattice system, and on-site solitons in the uniform
2D lattice described by the single DNLS equation, which corresponds to Eqs. (%
\ref{eq2}) with $\varepsilon =0$ \cite{pgk2d,nashi2d}. Along the line of $%
\varepsilon =0$ in Figs. \ref{fig2}, \ref{fig3} and \ref{fig5}, which
correspond to $\mu =-5$, one finds a stable symmetric complex, a stable
asymmetric mode (with one zero component), and stable antisymmetric
complexes. In terms of the uniform 2D lattice, the symmetric complex is
formed of two identical on-site-centered discrete solitons, which are stable
in the usual DNLS lattice \cite{pgk2d,nashi2d} [see, e.g., Fig. 5(a) in Ref.
\cite{nashi2d}]. As mentioned above, the introduction of the inter-lattice
linkage ($\varepsilon >0$) leads to the onset of the exponential instability
in the complex formed by two identical fundamental on-site solitons. The
symmetric complex recovers its stability at $\varepsilon \geq \varepsilon
_{c}$. Therefore, one can associate two bifurcation points with values $%
\varepsilon =0$ and $\varepsilon =\varepsilon _{c}$. The latter one was
actually identified above as the supercritical pitchfork bifurcation at
which the stable asymmetric branches disappear and the symmetric one is
re-stabilized.

\section{Conclusion}

In this work, we have introduced the 2D double-lattice nonlinear system,
linked in the transverse direction at a single site. The system can be
realized in terms of BEC, and may actually occur in a range of artificially
built discrete nonlinear media. The on-site nonlinearity was considered to
be self-focusing (the self-defocusing can be easily transformed the same
system by means of the staggering transformation \cite{PGK}). We have used
the VA (variational approximation) and numerical methods to find the regions
of existence, in the parameter plane of $(\mu ,\varepsilon )$ (the
propagation constant and the strength of the linkage between the lattices),
of the localized symmetric, asymmetric and antisymmetric soliton complexes
pinned to the linkage site. It was shown, by means of both approaches, that
the existence regions of the symmetric and asymmetric complexes are bounded.
The SSB (spontaneously symmetry-breaking)\ pitchfork bifurcation of the
supercritical part has been found, which destabilizes the symmetric
complexes and simultaneously creates stable asymmetric ones. The stability
of the antisymmetric complexes changes twice. Areas of the bistability
between the antisymmetric modes and either symmetric or asymmetric ones have
been found too. Direct simulations demonstrate that unstable symmetric modes
relax into the breathing asymmetric complexes, while the antisymmetric
solitons with large amplitudes are robust against perturbations,
transforming into strongly pinned breathers, which keep the antisymmetric
structure.

This work also suggests a possibility to create and investigate vortex
complexes in 2D parallel-coupled lattice systems, which will be reported
elsewhere.

\acknowledgments M. D. P., G. G., A. M., and Lj. H. acknowledge support from
the Ministry of Education and Science, Serbia (Project III45010).

\end{document}